\documentclass[preprint]{aastex}
\usepackage{epsf}
\usepackage{emulateapj5}
\usepackage{onecolfloat}
\usepackage{apjfonts}
\usepackage{amsmath}

\def\gtsim {\lower .1ex\hbox{\rlap{\raise .6ex\hbox{\hskip .3ex
        {\ifmmode{\scriptscriptstyle >}\else
                {$\scriptscriptstyle >$}\fi}}}
        \kern -.4ex{\ifmmode{\scriptscriptstyle \sim}\else
                {$\scriptscriptstyle\sim$}\fi}}}

\newcommand{\be}{\begin{equation}}
\newcommand{\eq}{\end{equation}}
\newcommand{\Mo}{{\rm M_\odot}}
\newcommand{\Moyr}{{\rm M_{\odot}yr^{-1}}}
\newcommand{\kms}{\>{\rm km}\,{\rm s}^{-1}}
\def\degrees{^\circ}
\def\LCDM{$\Lambda$CDM}

\def\kpc{\ {\rm kpc}}
\def\pc{\ {\rm pc}}

\begin{document}
\submitted{The Astrophysical Journal, accepted}
\vspace{1mm}
\slugcomment{{\em The Astrophysical Journal, accepted}}

\shortauthors{KAZANTZIDIS ET AL.}
\twocolumn[
\lefthead{Fate of Supermassive Black Holes and Evolution of the 
$M_{\rm BH}$--$\sigma$ Relation in Merging Galaxies}
\righthead{KAZANTZIDIS ET AL.}

\title{Fate of Supermassive Black Holes and Evolution of the 
$M_{\rm BH}$--$\sigma$ Relation in Merging Galaxies: 
The Effect of Gaseous Dissipation}

\author{Stelios Kazantzidis,\altaffilmark{1}
	Lucio Mayer,\altaffilmark{1}	
	Monica Colpi,\altaffilmark{2}
	Piero Madau,\altaffilmark{3}
	Victor~P. Debattista,\altaffilmark{4}
	James Wadsley,\altaffilmark{5}	
	Joachim Stadel,\altaffilmark{1} 
                Thomas Quinn,\altaffilmark{6}
	and Ben Moore\altaffilmark{1}
}

\begin{abstract}

We analyze the effect of dissipation on the orbital evolution
 of supermassive black holes (SMBHs) using high-resolution self-consistent
 gasdynamical simulations of binary equal- and unequal-mass mergers
 of disk galaxies. The galaxy models are consistent with the {\LCDM}
 paradigm of structure formation and the simulations include the 
 effects of radiative cooling and star formation. 
 We find that equal-mass mergers {\it always} lead to the 
 formation of a {\it close} SMBH pair at the center of the remnant
 with separations limited solely by the adopted force resolution of 
 $\sim 100\pc$. Instead, the final SMBH separation
 in unequal-mass mergers depends sensitively on how the central structure 
 of the merging galaxies is modified by dissipation. In the absence of dissipation,
 the satellite galaxy can be entirely disrupted before the merger is completed
 leaving its SMBH wandering at a distance too far from the
 center of the remnant for the formation of a close pair.
 In contrast, we show that gas cooling facilitates the pairing process by increasing the resilience 
 of the companion galaxy to tidal disruption. Moreover, we demonstrate that merging disk 
 galaxies constructed to obey the $M_{\rm BH}$--$\sigma$ relation, 
 move relative to it depending on whether they undergo a dissipational or 
 collisionless merger, regardless of the mass ratio of the merging systems. 
 Collisionless simulations reveal that remnants tend to move away from the mean 
 relation highlighting the role of gas-poor mergers as a possible source of scatter.
 In dissipational mergers, the interplay between strong gas inflows associated with the formation of 
 massive nuclear disks and the 
 consumption of gas by star formation provides the necessary fuel to the SMBHs and allows 
 the merger remnants to satisfy the relation.

\end{abstract}

\keywords{black hole physics --- cosmology: theory --- galaxies: mergers
--- hydrodynamics --- methods: numerical}
]

\altaffiltext{1}{Institute for Theoretical Physics, University of Z\"urich,
Winterthurerstrasse 190, CH-8057 Z\"urich, Switzerland; stelios@physik.unizh.ch.}
\altaffiltext{2}{Dipartimento di Fisica G. Occhialini, Universit\`a di
Milano Bicocca, Piazza della Scienza 3. I-20126 Milano, Italy.}
\altaffiltext{3}{Department of Physics, University of California at Santa Cruz,
1156 High Street, Santa Cruz, CA 95064.}
\altaffiltext{4}{Institut f\"ur Astronomie, ETH Z\"urich, Scheuchzerstrasse 7, 
CH-8093 Z\"urich, Switzerland.}
\altaffiltext{5}{Department of Physics and Astronomy, McMaster University, 
1280 Main Street West, Hamilton, ON L8S 4M1, Canada.}
\altaffiltext{6}{Department of Astronomy, University of Washington, Seattle, WA 98195.}

\section{INTRODUCTION}
\label{section:introduction}

Irrefutable dynamical evidence indicates that supermassive black holes 
(SMBHs) with masses ranging from $10^{6}$ to above $10^{9}\Mo$
reside at the centers of most galaxies hosting spheroids 
\citep[e.g.,][]{kormendy_richstone95}. The available data show the 
existence of a remarkably tight correlation between the mass of the 
SMBHs, $M_{\rm BH}$, and the stellar velocity dispersion of the host galaxy 
spheroidal component, $\sigma$ \citep{ferrarese_merritt00,gebhardt_etal00}, 
suggesting a fundamental mechanism that connects SMBH 
assembly and galaxy formation \citep[e.g.,][]{silk_rees98,burkert_silk01}. 

According to the currently favored cold dark matter cosmological models,  
structures in the universe are the end result of a complex hierarchy of
mergers and accretion of smaller subunits. Thus, the 
hierarchical buildup of SMBHs by massive seed black holes present at the 
center of protogalaxies and the formation of SMBH binaries appear as 
a natural consequence in any ``bottom-up'' cosmogony. Recently,
the discovery of a binary active galactic nucleus in the interacting system 
NGC 6240 by \citet{komossa_etal03} has lent support to this picture.
Although substantial gas accretion is also required for the SMBHs to reach
their present-day mass density and satisfy the $M_{\rm BH}$--$\sigma$ relation 
\citep{volonteri_etal03}, the link between these two main modes of SMBH 
growth is still missing.

The formation and subsequent orbital evolution of a SMBH pair 
depends on how efficiently the host galactic cores lose angular momentum 
by dynamical friction during the merging process. Once the two cores have 
merged, the SMBHs may decay further and eventually form a binary owing to the drag 
exerted by the background mass distribution. 
These two evolutionary phases have been investigated numerically by 
a number of authors \citep{governato_etal94,makino_ebisuzaki96,milosavljevic_merritt01,
makino_funato04}. However, the galaxy models used in these studies were 
idealized, spherical stellar systems that could at most faithfully represent the 
central spheroidal components of real galaxies. Hence, both the larger scale
dynamical evolution of the merging systems and the cosmological framework 
were missing. Moreover, these studies did not explore the role of a 
dissipative component in driving the evolution of a SMBH pair.
Notable exceptions are the studies by \citet{escala_etal04a,escala_etal04b}, 
who showed that gas causes continuing loss of angular momentum 
and rapidly reduces the relative SMBH separation to distances where gravitational 
radiation and coalescence is efficient. However, if and how their initial 
conditions are related to the larger scale dynamics involved in galaxy 
merging is still unclear.

In this Letter we report on the effects of gaseous dissipation on the
fate of SMBHs using high-resolution binary disk galaxy merger simulations.
\citet{governato_etal94} have already highlighted the difficulty of
forming a close SMBH pair when tidal disruption of one of the host systems
intervenes. As we illustrate below, gaseous dissipation is vital for the survival of the 
host galaxies by deepening their potential wells and provides the necessary fuel 
for the SMBH growth. Finally, we analyze the stellar kinematics and distribution of gas 
in the central regions of the merger remnants to investigate for the first time how 
merging galaxies move with respect to their initial location on the $M_{\rm BH}$--$\sigma$ 
plane.

\begin{figure*}[t]
\centerline{\epsfxsize=12.8cm \epsffile{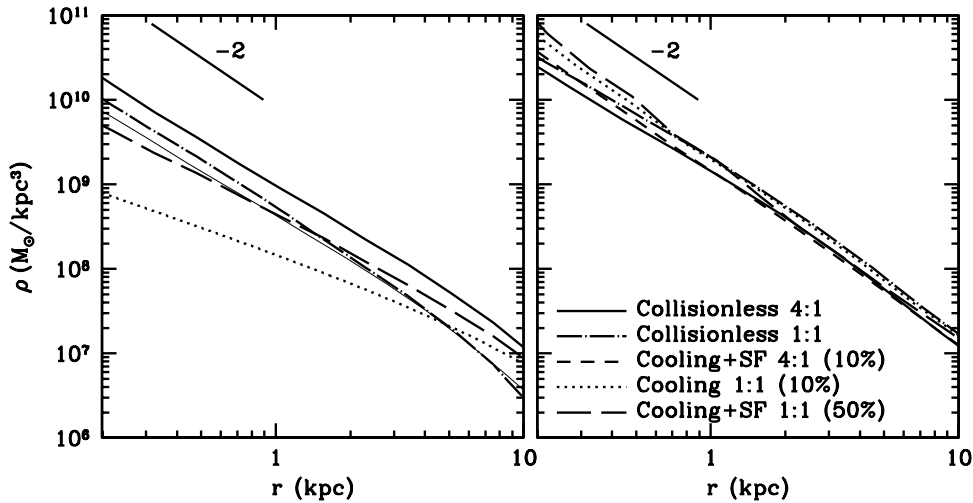}}
\caption{Spherically averaged density distribution for different components in the initial
models ({\it left}) and some of the remnants ({\it right}) plotted from the 
force resolution outward. {\it Left:} Different thick curves show the 
total ({\it solid}), dark matter density after the adiabatic contraction due to 
baryons ({\it dashed}) and total density of baryons ({\it dot-dashed}) in the basic galaxy model. The 
dotted curve shows the 
dark matter density before the baryonic contraction. The thin solid curve shows the 
{\it total} density of the satellite galaxy. After the adiabatic contraction the central 
density slope is close to isothermal with $\rho(r)\propto r^{-2}$ for both 
galaxy models. {\it Right:} Total density profiles for some of the merger remnants in the 
coplanar encounter geometry. The numbers in brackets indicate the values for the gas 
fraction, $f_{\rm g}$, used in the particular simulation. The center of each remnant was 
defined as the location of the minimum of the gravitational potential. In dissipative 
simulations, the remnants exhibit an increase in the central density resulting from strong 
gas inflows and star formation associated with the merger.}
\label{fig1}
\end{figure*}
%

\section{NUMERICAL SIMULATIONS}
\label{section:num_sim}

We perform four types of high-resolution simulations of binary disk galaxy
mergers with mass ratios of $1$:$1$ and $4$:$1$ including different 
physical processes. We simulate purely collisionless 
mergers and mergers in which we follow the gas dynamics 
``adiabatically,'' i.e., without radiative cooling. A third set of simulations
includes radiative cooling, while in the fourth set we also 
allow the cold gas to form stars. 
The simulations were performed with GASOLINE, a multi-stepping, parallel 
TreeSPH $N$-body code \citep{wadsley_etal04}.
We include radiative and Compton cooling for a primordial 
mixture of hydrogen and helium. The star formation algorithm is based on that 
by \citet{katz92}, where gas particles in dense, cold Jeans unstable regions and in 
convergent flows spawn star particles at a rate proportional to the local dynamical
time \citep[see also][]{governato_etal04}, and reproduces the Schmidt law. 
The star formation efficiency was set to $0.1$
which yields a star formation rate of $1-2\, \Moyr$ for models in isolation that have a 
disk gas mass and surface density comparable to those of the Milky Way.

The multicomponent galaxy models are constructed using the technique described 
by \citet{hernquist93} and their structural parameters are consistent with 
the {\LCDM} paradigm for structure formation \citep{mo_etal98}.
To this end, each galaxy consists of a spherical and isotropic \citet{navarro_etal96} 
dark matter (DM) halo \citep{kazantzidis_etal04a}, an exponential disk, and a spherical, 
\citet{hernquist90} non-rotating bulge.
For the basic galaxy model we adopted parameters 
from the Milky Way model A1 of \cite{klypin_etal02}. Specifically, the DM
halo had a virial mass of $M_{\rm vir}=10^{12}\Mo$, a
concentration parameter of $c=12$, and a dimensionless spin parameter
of $\lambda=0.031$. The mass, thickness and resulting scale length of the disk
were $M_{\rm d}=0.04 M_{\rm vir}$, $z_{0}=0.1 R_{\rm d}$, and $R_{\rm d}=3.5\kpc$, 
respectively. The bulge mass and scale radius were $M_{\rm b}=0.008 M_{\rm vir}$
and $a=0.2 R_{\rm d}$, respectively. The DM halo was adiabatically contracted 
to respond to the growth of the disk and bulge \citep{blumenthal_etal86} resulting
into galaxy models with a central total density slope close to isothermal
(Figure~\ref{fig1}). The companion galaxy is either a replica of the same model in 
equal-mass mergers or a system containing one-fourth of the mass in each 
component in which lengths and velocities are renormalized according to the 
cosmological scaling for virialized systems \citep{mo_etal98}. The disk scale length 
of the satellite galaxy is equal to $R_{\rm d}=2.2\kpc$.
Given the uncertainties in the adopted M/L, our galaxies are consistent
with the stellar mass Tully-Fisher and size-mass relations.
To each of the galaxy models we added a particle representing a SMBH 
at the center of the bulge component that was initially at rest.
For the larger galaxy model we used a SMBH mass equal to 
$M_{\rm BH}=3 \times 10^{6} M_{\odot}$.

We considered two values for the gas fraction, $f_{\rm g}$, namely
10\% and 50\% of the total disk mass. We shut off radiative 
cooling at temperatures below $2 \times 10^{4}$~K that is 
about a factor of $2$ higher than the temperature at which atomic radiative 
cooling would drop sharply due to the adopted cooling function. 
With this choice we take into account non-thermal pressures and approximate 
the simulated gas with the warm ISM of a real galaxy \citep{barnes02}. 
Moreover, this choice enables us to simulate very gas-rich disks that would
otherwise become strongly gravitationally unstable and undergo widespread
fragmentation. Two main encounter geometries were adopted: prograde coplanar mergers 
and mergers in which one of the disks was inclined with respect to the orbital plane.
We also simulated a single retrograde equal-mass merger in which both disk spin 
vectors were antiparallel to the orbital angular momentum vector.
The galaxies approached each other on parabolic orbits with pericentric distances 
that were 20\% of the more massive galaxy's virial radius, typical of cosmological 
mergers \citep{khochfar_burkert03}. The initial separation of the halo centers was twice 
their virial radii and their initial relative velocity was determined from the corresponding 
Keplerian orbit of two point masses. In collisionless mergers, we used  
$N=1.2 \times 10^6$ particles in total, with each galaxy consisting of $10^5$ 
disk particles, $10^5$ bulge particles, and $10^6$ DM particles. 
The gas component in gasdynamical simulations was represented by $10^5$ particles.
We adopted a gravitational softening of $\epsilon = 0.1\kpc$
for both the DM and baryonic particles of the larger galaxy, and 
for its SMBH $\epsilon=0.03\kpc$, 
which is small enough to follow its orbital evolution.
For the satellite galaxy, all softening lengths were scaled according to 
$\epsilon \propto m^{1/3}$. In all simulations the merger remnants were allowed to 
settle into equilibrium $\sim 15$ dynamical times at the disk half-mass radius 
after the merger was completed.

\section{RESULTS}
\label{section:results}

\subsection {SMBH Pairing in Binary Disk Galaxy Mergers}
\begin{figure}[t]
\centerline{\epsfxsize=3.2in \epsffile{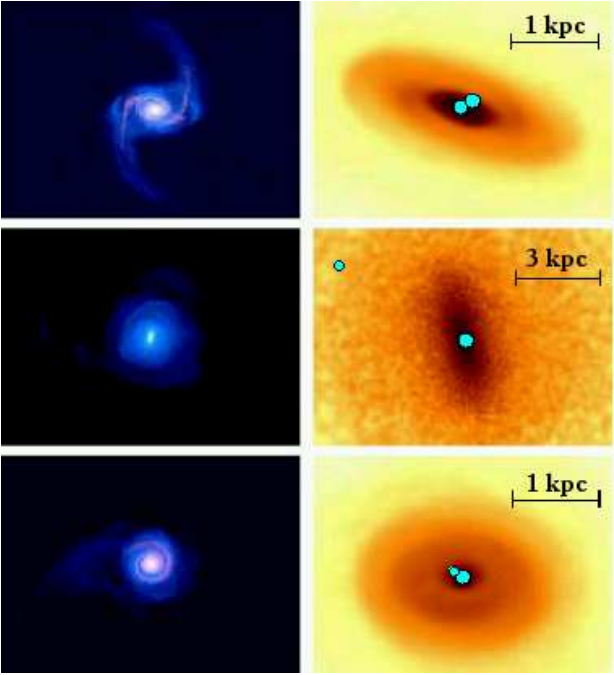}}
\caption{Final separation of SMBHs ({\it filled circles}) in a subset 
of merger simulations. The large scale ({\it left}) and small-scale ({\it right})
structure of the remnants projected onto the orbital plane is also shown.
All frames correspond to remnants that were allowed to relax for several 
dynamical times after the merger was complete.
The top and bottom rows of panels present results for the 
coplanar $1$:$1$ and $4$:$1$ mergers, respectively, with gas cooling.
The middle rows of panels corresponds to the coplanar collisionless 
$4$:$1$ merger. The frames on the left show the logarithmic baryonic 
surface density maps and are $320 \times 230\kpc$.
The limiting surface density is $\Sigma=1\Mo/\pc^{2}$. 
Blue and red maps are used for the stellar and gaseous component, respectively,
and adaptive smoothing is used to preserve details in 
high-density regions. The top and bottom frames on the right show the central 
gaseous disks, while the middle frame on the right shows the 
stellar distribution of the merger remnant. 
\label{fig2}}
\end{figure}
The galaxies merge in three to five orbits (between $5.5$ to $7$ Gyr) 
depending on the mass of the companion, a timescale that is 
set by dynamical friction and tidal stripping \citep[e.g.,][]{taffoni_etal03}.
These timescales are longer than those reported in earlier studies
\citep[e.g.,][]{barnes_hernquist96} due to the larger and more realistic 
pericentric distances adopted here.

In equal-mass mergers, the cuspy potentials of both galaxies are deep enough to 
allow the survival of their inner regions until orbital decay by dynamical friction 
is complete. This result holds for both collisionless and gasdynamical mergers.
In the latter, during the first two orbits a strong spiral pattern appears in both the stellar and the
gaseous component and mild non-axisymmetric torques redistribute mass and 
angular momentum driving approximately 20\% of the gas towards the center. 
However, the central bulges stabilize the galaxies against bar formation. 
Shortly before the galaxies merge, a second much stronger gas inflow
occurs caused by strong tidal torquing as the galaxies perform a close fly-by.
However, only in the simulations with radiative cooling does this inflow result 
in a significant central concentration of cold ($T \sim 2 \times 10^{4}$~K) gas.
In this case, more than 80\% of the gas that was originally in the disks is 
collected within the central $500\pc$. This results in a considerable
steepening of the central density slope of the combined distribution of DM 
and baryons (Figure~\ref{fig1}). In adiabatic simulations, only 20-30\% 
of the initial gas mass collects within the central $\kpc$. This gas is heated  
to temperatures $T >10^5$~K by shocks and compressions during the merger and 
subsequently expands since it cannot radiate away its energy.

Large central gas inflows were already noticed in lower
resolution merger simulations \citep[e.g.][]{barnes_hernquist96,barnes02}.
In simulations with star formation these large inflows result in a central 
starburst and more than 90\% of the central gas distribution is converted into stars in less 
than $10^8$~yr. The peak star formation rates range from $30$ to
$>100\ \Moyr$, comparable to those of luminous infrared galaxies
(LIRGs) and ultraluminous infrared galaxies (ULIRGs), suggesting that
our modeling of star formation and the amount of cold gas present 
are realistic. The two SMBHs end up orbiting at the center of the 
remnant on eccentric orbits and form a {\it close} pair
at a separation comparable to the adopted force resolution of $\sim 100\pc$ (Figure~\ref{fig2}).
Significantly higher mass and force resolution would be needed to follow 
the SMBHs at relative separations where they would form a binary system.

In unequal-mass mergers, the outcome depends sensitively on how the 
internal structure of the merging galaxies is modified by dissipation.
In collisionless simulations, the tidal disruption of the satellite galaxy
at about $6$ kpc from the center leaves its SMBH wandering at a distance 
that prohibits the formation of a close pair. This ``naked'' SMBH would contribute to 
a population of wandering SMBHs in galactic halos \citep{volonteri_etal03}
At such a distance the timescale for orbital decay predicted
by dynamical friction would exceed a Hubble time.
In simulations with cooling, the situation is completely reversed as the gas inflow becomes
particularly strong in the companion galaxy. We measured an inflow of about 
$3\ \Moyr$ within the central $\sim$~kpc compared with $<0.5\ \Moyr$ for 
the larger galaxy. This striking difference in the gas inflows
is attributed to a tidally induced strong stellar bar observed in the satellite galaxy 
approximately $1.5$ Gyr before the final passage.
More than $50\%$ of the gas is funneled into the center owing to gas shocking in 
the bar potential. At this stage the companion galaxy has about 20\% more gas within its 
central $\sim$~kpc relative to the larger galaxy.
Dissipation enables the core of the infalling galaxy to survive complete tidal 
disruption by deepening its potential well resulting in a
SMBH pair separated by $\sim 100\pc$ as in equal-mass mergers (Figure~\ref{fig2}).

The large inflows observed in the cooling and star formation simulations always produce a 
rotationally supported nuclear disk with a size in the range $1-2\kpc$
and a temperature set by our temperature floor ($T = 2 \times 10^{4}$~K). 
In adiabatic simulations, this gas remains in a hot phase and forms a pressure 
supported cloud of similar size. The nuclear disks result from the 
coalescence of two central disk-like structures formed at the centers of the 
galaxies primarily by the strong inflow just before they merge.
These disks are tilted by several to a few tens of degrees relative to the orbital plane of 
the galaxies. They have peak rotational velocities and gas masses in the range of
$250-300\kms$ and $\sim 10^8-10^9 M_{\odot}$, respectively, and they
are well-resolved ($N \gtsim10^4$) in our high mass resolution simulations.
Nuclear disks of {\it molecular} gas of a few hundred pc to a 
$\sim$~kpc scale have been identified spectroscopically for several AGNs and 
ULIRGs \citep[e.g.][]{downes_solomon98,davies_etal04}.
The observed sizes, masses, and rotational velocities are comparable to 
those measured in our simulations. The nuclear disks reported here 
provide the reservoir of gas that fuels the central SMBHs. 

\subsection {$M_{\rm BH}$--$\sigma$ Relation in Binary Disk Galaxy Mergers}

For each remnant, we calculate line-of-sight aperture stellar velocity 
dispersions $\sigma$ within one effective radius $R_{\rm e}$ \citep{gebhardt_etal00} and we
use a range of viewing orientations, namely inclinations $i=0$, $60\degrees$, and 
$90\degrees$. For $i=60\degrees$ and $i = 90\degrees$, we rotate the system about an
axis perpendicular to the inclination axis through an angle $\phi$ set to 
0, $45\degrees$, $90\degrees$, and $135\degrees$. 
Thus, for each simulation we obtain $9$ different orientations. 
For each of these viewing angles, we measured the mass density profiles in 
circular apertures out to two radii ($8$ and $25\kpc$) under two assumptions: 
(1) the system is a disk galaxy and decompose the 
mass density profile into a S\'ersic bulge and an exponential disk;
(2) the system is a pure spheroid and fit it with a pure S\'ersic profile.
This procedure ensures that the measurements of $R_{\rm e}$, and therefore 
of $\sigma$, that we obtain are very robust.

\begin{figure}[t]
\centerline{\epsfxsize=3.2in \epsffile{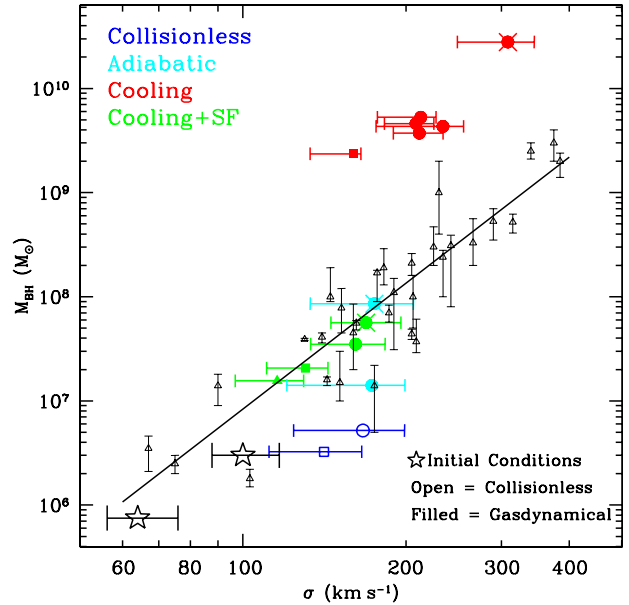}}
\caption{$M_{\rm BH}$--$\sigma$ relation in binary disk galaxy mergers.
The {\it open triangles} show data from the galaxy sample compiled by 
\citet{tremaine_etal02} and the {\it solid line} corresponds to their best-fit correlation. 
Results for simulated remnants are shown in color and
the initial galaxy models are denoted by {\it stars}.
Open and filled symbols correspond to collisionless and gasdynamical simulations,
respectively. {\it Circles} and {\it squares} show equal- and unequal-mass 
mergers, respectively. The {\it filled triangle} corresponds to an
unequal-mass merger in which the larger disk was inclined by $45\degrees$
with respect to the orbital plane. Symbols outlined with {\it crosses} correspond
to mergers with 50\% gas fraction in the disks. The error bars show the spread 
about the mean value of $\sigma$ in each remnant.}
\label{fig3}
\end{figure}

In cases when the two SMBHs form a close pair, the $M_{\rm BH}$ 
assigned to each remnant is calculated by summing the masses of the two SMBHs 
and adding the gas mass contained within the resolution of our gasdynamical
simulations. In doing this we implicitly assume that any close pair will actually 
form a SMBH binary that will subsequently coalesce.
Indeed, this is supported by the smaller scale simulations of 
\citet{escala_etal04a,escala_etal04b}, who 
showed that SMBH pairs starting at a separation of few hundred pc 
in a gaseous background will typically merge 
within $\sim 10^7$ yr. When the SMBHs do not form a close pair, we only 
consider the mass of the SMBH located closer to the center of the remnant. 
Figure~\ref{fig3} shows the location of all merger remnants on the 
$M_{\rm BH}$--$\sigma$ plane together with real galaxy data and their best-fit correlation
taken from \citet{tremaine_etal02}.  We note that the initial galaxy models start very 
close to the mean correlation by construction. The error bars show the spread about the mean 
value of $\sigma$ and are smaller in the cooling and star formation
simulations likely due to the fact that these remnants are more spherical \citep{kazantzidis_etal04b}. 
Mergers lead to an increase of the stellar velocity dispersion that is 
remarkably different depending on the physics included in the simulations. The 
largest increase occurs in simulations with radiative 
cooling which exhibit the deepest potentials. When star formation is 
included, we measure velocity dispersions closer to that of the collisionless
simulations, since the gas does not build a significant central mass concentration.

Figure~\ref{fig3} illustrates that collisionless mergers move remnants away from
the mean $M_{\rm BH}$--$\sigma$ relation. In adiabatic simulations, the remnants
have in principle enough gaseous fuel to remain close to the relation, but the gas is too hot 
to accrete onto the SMBHs. In mergers with radiative cooling, the total amount of 
nuclear gas is more than one order of magnitude larger than needed to keep the galaxies 
close to the relation. This should also be regarded as an upper limit on the gas 
mass that may accrete onto the SMBHs since feedback processes ensuing once the 
AGN becomes active will regulate the inflow and allow the accretion of 
only a fraction of this gas, thus limiting the SMBH growth \citep{silk_rees98}.
If AGN feedback is strong enough to supress cooling a hot gas phase
similar to that found in our adiabatic simulations is expected to form. Thus,  
the results from cooling and adiabatic simulations likely provide 
upper and lower bounds to the true physical behavior of the resulting remnants.
Note that the point corresponding to the $4$:$1$ merger is the highest with respect to the 
best-fit correlation in Figure~\ref{fig3}. This is the product of the particularly strong gas 
inflow occurring in the satellite galaxy just before the merger is completed. 
Overall a higher fraction of the available gas is driven towards the center relative to the 
equal-mass galaxy mergers.  When star formation is included the amount 
of cold gas in the nuclear disks is reduced by more than 90\% and the merger remnants
move nearly parallel to the relation. This suggests that star formation
during mergers is a key ingredient for maintaining the tightness 
of the $M_{\rm BH}$--$\sigma$ relation. 

\section{DISCUSSION AND CONCLUSIONS}
\label{section:discussion_conclusions}

Gaseous dissipation influences considerably the outcome of 
binary mergers of disk galaxies containing SMBHs. Most importantly, 
it controls the SMBH pairing process in unequal-mass mergers by 
modifying the central structure of the companion galaxy, enabling it
to survive complete tidal disruption. Dissipationless, unequal-mass mergers
are also expected to produce close SMBH pairs when the satellite galaxy is dominated 
by a dense stellar central component, as in nucleated dwarf elliptical (e.g. M32) and 
S0 galaxies. Our results suggest that semi-analytic 
models of hierarchical SMBH growth that neglect the effect of dissipation 
on the orbital evolution of SMBHs underestimate their pairing efficiency
and subsequent coalescence. This consequently leads into 
overestimating the number of wandering SMBHs in MW-sized galaxies 
\citep{volonteri_etal03}. The higher pairing efficiency of $\sim 10^6 M_{\odot}$ 
SMBHs reported here has important implications for the probability of observing 
coalescence events with space-based gravitational waves experiments such as 
{\sc LISA}\citep{sesana_etal04}.

Gaseous dissipation also determines how merging galaxies constructed to satisfy 
the $M_{\rm BH}$--$\sigma$ relation move relative to their initial location on 
the $M_{\rm BH}$--$\sigma$ plane.  The collisionless simulations reveal
that when galaxies become gas poor which is likely at low $z$, their merger remnants 
will tend to move away from the mean relation. This suggests that gas-poor 
mergers act as a possible source of scatter in the mean $M_{\rm BH}$--$\sigma$ 
relation. The natural scaling between 
the amount of cold nuclear gas and the increase in the central stellar velocity dispersion 
explains why merger remnants in simulations with radiative cooling 
move above the mean relation. Star formation serves to move merger remnants nearly parallel 
to the relation. The fact that unequal-mass mergers
appear effective at building a reservoir of gas for SMBH accretion is noteworthy,
since in hierarchical structure formation models they are significantly more frequent 
than equal-mass ones.

The nuclear disks at the center of the remnants show significant spiral
patterns (Figure~\ref{fig2}) which transport angular momentum outwards and sustain 
radial inflows towards the center. We measure typical inflows in the 
range $10^{-2}-10^{-3}\ \Moyr$ within $1\kpc$. The spiral instabilities 
are sustained by the self-gravity of the disks. In star formation simulations
which contain the lightest disks, the minimum Toomre Q parameter \citep{toomre64}
is $\sim 2$ within the same distance. Since gravity is softened at scales 
corresponding to a significant fraction of the disk size ($200\pc$), short wavelength modes
are stabilized and the disks behave effectively as if they had
a higher Q parameter \citep{mayer_etal04}. Therefore, spiral modes should be
more vigorous with increasing resolution and promote even stronger inflows. 
Such inflows could feed the SMBHs and bridge the gap between the 
large scale flows and the viscous accretion taking place once the gas has 
reached the AGN accretion disk at subparsec scales \citep{heller_shlosman94,fukuda_etal00}.
Future simulations will be used to explore the feasibility of such mechanism.
Interestingly, observed nuclear disks also have Q values between 
$1$ and $2$ and some of them exhibit significant non-axisymmetric structure 
\citep{downes_solomon98}. 

\acknowledgments

S. K. would like to thank Aaron Dutton, Andrey Kravtsov and David Merritt for 
useful discussions. P. M. acknowledges support by NASA grants NAG5-11513 and
NNG04GK85G, and by NSF grant AST-0205738. The numerical simulations 
were performed on the zBox supercomputer at the University of Z\"urich
and on the Intel cluster at the Cineca Supercomputing Center
in Bologna (Italy).

\end{document}